\documentclass[prl,twocolumn,preprintnumbers,superscriptaddress,floatfix]{revtex4}

\usepackage{amsmath}
\usepackage{graphicx}
\usepackage[small]{subfigure}
\usepackage[hyperfootnotes=false]{hyperref}
\usepackage{braket}
\usepackage{cancel}
\usepackage{xspace}

\usepackage{color}
\definecolor{darkgreen}{rgb}{0,0.5,0}

\newcommand{\OMIT}[1]{}

\newcommand{\nn}{\nonumber}

\newcommand{\Tau}{\mathcal{T}}
\newcommand{\ttwoone}{\tau_{21}}
\newcommand{\hemi}{\text{hemi}}

\allowdisplaybreaks[4]

\DeclareRobustCommand{\Fig}[1]{Fig.~\ref{#1}}

\newcommand{\be}{\begin{equation}}
\newcommand{\ee}{\end{equation}}

\def\cT{\mathcal{T}}
\def\hcT{\widehat\mathcal{T}}

\def\dg{\dagger}

\newcommand{\pythia}{{\sc Pythia}\xspace}

\begin{document}


\preprint{\vbox{\hbox{MIT--CTP 4353}\hbox{\today}}}

\title{\boldmath Precision Jet Substructure from Boosted Event Shapes
\vspace{0.2cm}
}

\author{Ilya Feige}
\affiliation{Center for the Fundamental Laws of Nature,
Harvard University, Cambridge, Massachusetts 02138, USA}

\author{Matthew D. Schwartz}
\affiliation{Center for the Fundamental Laws of Nature,
Harvard University, Cambridge, Massachusetts 02138, USA}
  
\author{Iain W. Stewart}
\affiliation{Center for Theoretical Physics, 
  Massachusetts Institute of Technology, Cambridge, Massachusetts 02139, USA
}

\author{Jesse Thaler\vspace{0.2cm}}
\affiliation{Center for Theoretical Physics, 
  Massachusetts Institute of Technology, Cambridge, Massachusetts 02139, USA
}

\begin{abstract}

  Jet substructure has emerged as a critical tool for LHC searches, but studies
  so far have relied heavily on shower Monte Carlo simulations, which formally approximate
  QCD at leading-log level.  We demonstrate that systematic higher-order QCD
  computations of jet substructure can be carried out by boosting global event
  shapes by a large momentum $Q$, and accounting for effects due to finite jet
  size, initial-state radiation (ISR), and the underlying event (UE) as $1/Q$
  corrections.  In particular, we compute the 2-subjettiness substructure
  distribution for boosted $Z \to q \bar q$ events at the LHC at
  next-to-next-to-next-to-leading-log order.  The calculation is greatly
  simplified by recycling the known results for the thrust distribution in $e^+e^-$
  collisions.  The 2-subjettiness distribution quickly saturates, becoming $Q$
  independent for $Q\gtrsim 400$ GeV.  Crucially, the effects of jet
  contamination from ISR/UE can be subtracted out analytically at large $Q$,
  without knowing their detailed form.  Amusingly, the $Q=\infty$ and $Q=0$
  distributions are related by a scaling by $e$, up to next-to-leading-log
  order.

\end{abstract}

\maketitle


The Large Hadron Collider (LHC) is exploring a new regime where the collision
energy far exceeds the masses of known standard model particles.  At such energies, heavy
particles such as $W/Z$ bosons and top quarks are often produced with large
Lorentz boost factors, which leaves their hadronic decay products collimated
into a single energetic ``fat jet''. Jet substructure techniques extract
information from these fat jets to distinguish boosted heavy objects from the
QCD background of jets initiated by light quarks and gluons. Examples of
variables defined for this purpose include planar
flow~\cite{Thaler:2008ju,Almeida:2008yp}, jet
angularities~\cite{Almeida:2008yp}, pull~\cite{Gallicchio:2010sw},
$N$-subjettiness~\cite{Thaler:2010tr,Kim:2010uj}, dipolarity~\cite{Hook:2011cq},
and angular correlations~\cite{Jankowiak:2011qa}, with applications to boosted
Higgs bosons~\cite{Butterworth:2008iy}, tops~\cite{Kaplan:2008ie,Thaler:2008ju},
$W$s~\cite{Cui:2010km} and quark versus gluon
discrimination~\cite{Gallicchio:2011xq}, along with many beyond the standard
model applications (see~\cite{Abdesselam:2010pt,Altheimer:2012mn} for recent
reviews).  Jet substructure measurements are underway at the
LHC~\cite{CMS-PAS-EXO-11-006,Aad:2012ef}, but to date, studies of the analyzing
power of substructure variables have been limited by the use of leading-log
shower Monte Carlo simulations.  If higher-order QCD computations were
available, one could use them to directly compare to experiments or
test the accuracy of Monte Carlo simulations.

In this paper, we develop a framework for performing jet substructure
computations analytically, in the limit where the boosted object of interest has
a large momentum $Q$.  We find a mapping between global $e^+e^-$ event
shapes---which have been calculated to high precision---and jet
substructure variables in the large $Q$ limit, treating finite jet size, initial
state radiation (ISR), and underlying event (UE) as $1/Q$ corrections.
Concretely, we consider the jet substructure observable $N$-subjettiness
$\Tau_N$~\cite{Thaler:2010tr}, which is the subjet version of the global event
shape $N$-jettiness~\cite{Stewart:2010tn}.  The ratio $\Tau_N/\Tau_{N-1}$ is a
robust probe for $N$-prong decays~\cite{Thaler:2011gf}, and compares favorably
to other methods for boosted object identification.

Here, we focus on 1- and 2-subjettiness ($\cT_1$ and $\cT_2$), which are
relevant for LHC searches involving $W/Z$ and Higgs bosons.  We compute the
distribution for the ratio $\cT_2/\cT_1$ from $Z \to q \bar q$ decays to
next-to-next-to-next-to-leading-log (N$^3$LL) order, using ingredients from
higher-order calculations of the classic $e^+e^-$ thrust event
shape~\cite{Farhi:1977sg,Catani:1991kz,Korchemsky:1999kt,Fleming:2007qr,
  Schwartz:2007ib,Becher:2008cf,Abbate:2010xh}.  From a calculational point of
view, the use of this ratio is crucial, since it has a finite limit when $Q\to
\infty$.  We will show that our full subjet distribution is equal to the global
distribution generated by the $Z$ decay products, up to $1/Q$ power-suppressed
corrections.  The dominant hadronization corrections cause a shift which is
encoded in a single $Q$-independent parameter.
We compare our substructure calculation to \pythia~8.150
\cite{Sjostrand:2007gs} tune 4C and also use \pythia to demonstrate that the effects from
the jet boundary and from external radiation (i.e.\ ISR and UE) are suppressed by
$1/Q$, only entering at the 5\% level for $Q\gtrsim 400\,{\rm GeV}$.

We begin by considering a fat jet of size $R$ (clustered with anti-$k_T$
\cite{Cacciari:2008gp}) in a $pp$ collision event.  This jet should contain
 most of the $Z$ decay products as well as some ISR/UE contamination. 
The jet momentum is $P_J^\mu =\sum _{j \in J} p_j^\mu$, where $j$ runs over the
four-vectors $p_j^\mu$ within the jet $J$.  The jet boost $Q$ is defined as
$Q\equiv |\vec{P}_J|$.  To calculate $N$-(sub)jettiness, we must specify a
distance
measure~\cite{Stewart:2010tn,Thaler:2010tr,Jouttenus:2011wh,Thaler:2011gf}, and
we use the geometric measure 
\be \label{eq:Tn} 
\cT_N \equiv \min\limits_{n_1, n_2, \ldots, n_N}\sum\limits_{j\in J} 
 \min \{n_1 \cdot p_j , n_2 \cdot p_j, \ldots, n_N \cdot p_j \}.  
\ee
Here, $n_i^\mu=(1,\hat{n}_i)$ are lightlike axes defined by the overall
minimization.  The minimum inside the sum partitions the jet's constituents into
subjet regions $J_1,\ldots, J_N$, defined by the axes $n_i^\mu$. For the
$N$-jettiness event shape, $J$ is replaced by the entire event.

For 1-subjettiness, $\Tau_1 = \min\limits_{n}\sum_{j\in J} n \cdot p_j$, which
can also be written as the small component of the fat-jet momentum, $\Tau_1 =
P^+ \equiv n\cdot P_J$.  If the jet contained all the $Z$ decay products and
nothing else, $\Tau_1$ would depend only on the $Z$ boson momentum $P_Z^\mu$ as:
\begin{align} \label{eq:T12h}
 \hcT_1 \equiv P_Z^+ = \sqrt{Q^2+m_Z^2} -Q.
\end{align}
Thus, the  difference
\be \label{eq:deltaTau}
\Delta \tau \equiv \cT_1-\hcT_1
\ee
measures how much the $Z$ is incorrectly reconstructed. We will use $\Delta\tau$ to correct for ISR/UE contamination. 

\begin{figure}
\begin{center}
\includegraphics[width=0.5\columnwidth]{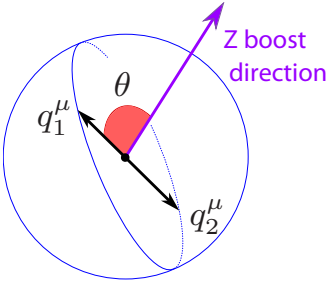}
\end{center}
\vspace{-0.4cm}
\caption{Kinematics of boosted $Z$ decay.  }
\label{fig:KinematicsZDecay}
\end{figure}

Turning to 2-subjettiness, we first calculate the ratio $\cT_2/\cT_1$ including
only the $Z$ decay products, and then discuss how other effects can be
systematically included. The distribution for the $Z$ decay products is easily
determined by boosting the $Z$ rest frame distribution.  At leading order, $Z$
decays to a $q\bar q$ pair which go off back-to-back in the rest frame, at an
angle $\theta$ (the {\it helicity angle}) with respect to the boost axis as in
\Fig{fig:KinematicsZDecay}. For simplicity, we treat the $Z$ as unpolarized with
a flat $\theta$ distribution, but one could easily integrate over a different
$\theta$ distribution, for example for $W$s coming from top
decays~\cite{Kaplan:2008ie}.  In the boosted frame, the $Z$ momentum $P_Z^\mu$
and the two daughter-quark momenta $q_1^\mu$ and $q_2^\mu$ are
\begin{align} \label{eq:kinematics1}
 P_Z^\mu &= \Big\{E_Q,0,0,Q\Big\},
 \\
 q_1^\mu &= \frac12\Big\{E_Q-Q\cos\theta,-m_Z\sin\theta,0,Q-\cos\theta E_Q\Big\}, 
 \nn\\
 q_2^\mu &=  \frac12\Big\{E_Q+Q\cos\theta,m_Z\sin\theta,0,Q+\cos\theta E_Q\Big\}
 , \nn
\end{align}
with $E_Q=\sqrt{m_Z^2+Q^2}$. The quark energies are  $E_1=\frac{1}{2}(E_Q - Q\cos\theta)$ and $E_2=\frac{1}{2}(E_Q + Q\cos\theta)$.

For the relevant small $\cT_2$ region, the subjet directions from the
minimization in \eqref{eq:Tn} can be aligned with the leading-order quark
directions~\cite{Stewart:2010tn}. Thus, we can take
\begin{align}
  n^\mu &= (1,0,0,1) \,, 
  & n_1^\mu &= \frac{1}{E_1}q_1^\mu \,,
  & n_2^\mu &= \frac{1}{E_2}q_2^\mu\, ,
\end{align} 
where $n^\mu$ is the $\cT_1$ axis  and  $n_1^\mu$ and $n_2^\mu$ are the $\cT_2$ axes.
In terms of the subjet masses $m_i$ and energies $E_i$,
\begin{align}
\cT_2 = P_1^+ + P_2^+ \simeq \frac{m_1^2}{2E_1} + \frac{m_2^2}{2E_2}.
\end{align}
In the large $Q$ limit,
$E_1\sim Q \sin^2(\theta/2)$, 
$E_2 \sim Q\cos^2(\theta/2)$, and
 $\cT_1 \sim m_Z^2/(2Q)$, while $m_i$ are $Q$ independent. Thus the distribution of the ratio $\cT_2/ \cT_1$
asymptotes to a fixed $Q$-independent result. 

Now let us consider how the scaling with $Q$ is affected when $\cT_2/\cT_1$ is
considered in a realistic environment, such as at the LHC.  A measurement of
$\cT_2/\cT_1$ includes effects from having a finite jet boundary and from
including radiation from elsewhere in the event.  The jet boundary $R$
identifies a $Q$-independent phase space region about the jet axis.  As $Q\to
\infty$, the phase space for the $Z$ decay products to land outside of the cone
falls as $1/Q$.  Hence, the jet boundary is at most a $1/Q$ correction to
$\cT_2/\cT_1$. The same conclusion holds if $R$ is defined with a
jet algorithm other than anti-$k_T$.

Next consider radiation not coming from the $Z$ decay (i.e.~ISR/UE).  Since $\Tau_N$ depends linearly on $p_j^\mu$ in \eqref{eq:Tn}, both
$\cT_1$ and $\cT_2$ will be distorted by (different) shifts due to this
contaminating radiation.  If we require the fat-jet mass to be close to $m_Z$, then the shifts will scale as 
$\cT_N$, giving at most an $\mathcal{O}(Q^0)$ distortion of $\cT_2/\cT_1$.  To turn this into a $1/Q$ distortion, note that the distribution of contaminating radiation is smooth over the fat jet, and at large $Q$,
\begin{align}
 n_{1,2}^\mu = n^\mu +\frac{m_Z}{Q} \Big\{-\cot\frac{\theta}{2},\tan\frac{\theta}{2} \Big\}
   \hat e_x^\mu + {\cal O}\Big(\frac{1}{Q^2}\Big),
\end{align} 
where $\hat e_x^\mu=(0,1,0,0)$.  Comparing $n \cdot p_j$ and
$\min\{n_1 \cdot p_j, n_2 \cdot p_j\}$, both $\cT_1$ and $\cT_2$ will be shifted in
the \emph{same} way up to $1/Q$ corrections.  Hence we can
remove the leading effect of contamination with $\Delta \tau$ from
\eqref{eq:deltaTau}, by defining
\begin{align} \label{eq:T12}
  \ttwoone \equiv\frac{\cT_2 - \Delta \tau}{\cT_1 -  \Delta \tau} .
\end{align}
$\ttwoone$ has two important properties: first, it is close to $\cT_2/\cT_1$
since $\ttwoone = \cT_2/\cT_1$ if only the exact $Z$ decay products are
included; second, it is insensitive to jet contamination up to $1/Q$
corrections.  It is crucial that the $\Delta \tau$ correction be made
experimentally on an event-by-event basis; if only the $\cT_2/\cT_1$
distribution is measured, then the contamination will {\it not} be a $1/Q$
correction. The subtraction can be improved further by replacing $\Delta \tau$
with $\Delta \tau' \equiv \Delta \tau (1-\frac{\pi}{2} m_Z/Q)$ in the numerator
of \eqref{eq:T12}; the additional factor accounts for the average fractional
difference between $\cT_2$ and $\cT_1$ for uncorrelated soft radiation.  The above logic is also appropriate for event pileup.

To compute the $\ttwoone$ spectrum at leading order in $1/Q$, we 
calculate $\cT_2/\cT_1$ assuming only the $Z$ decay products are included in the
fat jet. We then average over the angle $\theta$. Using the correspondence with
2-jettiness, the factorization formula for $\cT_2/\cT_1$
is~\cite{Stewart:2010tn}
\begin{align}
\label{eq:factorization}
& \frac{1}{\sigma_0}
 \frac{d\sigma}{d\ttwoone} = H \!\int\! \frac{d\cos\theta}{2} 
 \!\!\int\!\! ds_1ds_2dk_1dk_2\,  S(k_1,k_2,\{n_i\},\mu) 
 \nn\\
&~\times J(s_1,\mu) J(s_2,\mu)
 \,\delta\Big(\ttwoone \!-\! \frac{k_1 \!+\! k_2}{\widehat\Tau_1}
  \!-\! \frac{s_1E_2\!+\!s_2E_1}{2E_1E_2\widehat\Tau_1}\Big),
\end{align}
where $\sigma_0$ is the tree-level cross-section given by the $Z$ decay rate.
Here $H=H(m_Z,\mu)$, $J(s_i,\mu)$, and $S(k_1,k_2,\{n_i\},\mu)$ are respectively
the $Z \to q \bar{q}$ hard function, inclusive jet function, and 2-jettiness
soft function.  $H$ and $J$ are known at ${\cal
  O}(\alpha_s^2)$~\cite{Matsuura:1988sm,Becher:2006qw}.  For simplicity, we
consider the narrow width limit, neglecting $\mathcal{O}(\Gamma_Z / m_Z)$
corrections.  We also neglect non-singular corrections at
$\mathcal{O}(\alpha_s)$.  These contribute less than 5\% in the peak of the
$\ttwoone$ distribution and can be included following
\cite{Becher:2008cf,Abbate:2010xh}.

We now show that the 2-jettiness soft function $S$ can be related to the
hemisphere soft function $S_{\hemi}$---relevant for thrust and heavy jet
mass---which is known perturbatively to ${\cal
  O}(\alpha_s^2)$~\cite{Kelley:2011ng,Hornig:2011iu}.  The soft function is
\begin{align} \label{eq:softDef}
& S(k_1,k_2,n_1 \cdot n_2 ,\mu,\Lambda) 
 \equiv \frac1{N_c}\: \mbox{$\sum_{X_s}$}\:
 \delta(k_1-n_1\!\cdot\! P_s^1)
\nn \\
&~\times \delta(k_2-n_2\!\cdot\!  P_s^2) 
 \bra{0}\overline{Y}_{\!\!n_2}^T Y_{n_1} \ket{X_s}
 \bra{X_s}Y^\dg_{n_1}\overline{Y}^*_{\!\! n_2}\ket{0}
 ,
\end{align}
where the $Y$'s are light-like Wilson lines and $P_s^{1,2}$ are the momenta of
the subjets $J_{1,2}$ in the state $\ket{X_s}$.  Rotational invariance implies that
the subjet directions only appear in the combination $n_1\cdot n_2$, and the
argument $\Lambda \equiv \Lambda_{\rm QCD}$
is a reminder of nonperturbative corrections contained in $S$.  The hemisphere
case corresponds to $n_1\cdot n_2 = 2$, so that $S_{\hemi}(k_L,k_R,\mu,\Lambda)=
S(k_L,k_R,2,\mu,\Lambda)$.  From \eqref{eq:Tn}, the partitioning into regions of
$2$-subjettiness is invariant under a common rescaling of the subjet direction,
$n_1\to \beta n_1$ and $n_2\to \beta n_2$. So \eqref{eq:softDef} satisfies
\begin{equation}
\label{eq:scalingRelation}
S(k_1,k_2,n_1\!\cdot\! n_2,\mu,\Lambda) = \beta^2 S(\beta k_1,\beta k_2,\beta^2
n_1\!\cdot\! n_2,\mu,\Lambda). \nn
\end{equation}
Choosing
\begin{equation}
\beta=\beta_\theta= \sqrt{\frac{2}{n_1\cdot n_2}} = \frac{\sqrt{m_Z^2+Q^2 \sin^2\theta}}{m_Z},
\end{equation}
we  find
\begin{align} \label{eq:Srelation}
 &S(k_1,k_2,n_1 \cdot n_2,\mu,\Lambda) 
 = \beta_\theta^2\, S \left(\beta_\theta k_1,\beta_\theta k_2,2,\mu,\Lambda\right)
  \nn \\
 &\quad
 =  S_{\hemi}\left(k_1,k_2,\mu/\beta_\theta,\Lambda/\beta_\theta\right) \,,
\end{align}
where we have rescaled all dimensionful arguments by $\beta_\theta^{-1}$ and
used the fact that $S$ has mass dimension $-2$.  

When $k_i\gg \Lambda/\beta_\theta$, the leading nonperturbative correction to
$S_{\rm hemi}$ is equivalent to a
shift~\cite{Dokshitzer:1997ew,Lee:2006nr,Hoang:2007vb}, $k_i\to k_i -
\Phi/\beta_\theta$, where $\Phi\sim \Lambda$ is $Q$-independent. Since
$\Tau_2$ in \eqref{eq:Tn} is not identical to thrust for massive hadrons, we
cannot use the value found in \cite{Abbate:2010xh}.  All the objects in
\eqref{eq:factorization} have known renormalization group equations, so we can
sum large logarithms of $\ttwoone$ up to N${}^{3}$LL (with a Pad\'e
approximation for the small contribution of the four-loop cusp anomalous
dimension). Thus for $\ttwoone \gg 2\Lambda/(\widehat\Tau_1\beta_\theta )$ we
have
\begin{align} \label{eq:Ufact}
& \frac{1}{\sigma_0}\frac{d\sigma}{d\ttwoone} 
= \widehat\Tau_1^2 \!\int\! \frac{d\cos\theta}{2}  H(m_Z,\mu_H)
U_H(m_Z,\mu_H,\mu_J)
 \nn \\
&\times \!\!\int\!\! dz_s \, ds_1 ds_2 J\big(s_1,\mu_J\big)J\big(s_2,\mu_J\big)
S_\tau\Big(\widehat\Tau_1 z_s, \frac{\mu_S}{\beta_\theta},\alpha_s(\mu_S)\!\Big)
 \nn \\
&\times U_S^{\tau}\Big(\widehat\Tau_1 \ttwoone 
  \!-\! \frac{2\Phi}{\beta_\theta}
  \!-\!  \frac{s_1}{2E_1}\!-\!  \frac{s_2}{2E_2}
  \!-\!\widehat\Tau_1 z_s , \frac{\mu_J}{\beta_\theta}, \frac{\mu_S}{\beta_\theta}\Big).
\end{align}
Here $S_\tau$ is the perturbative thrust soft function, and $H$, $J$, and
$S_\tau$ are fixed-order expansions in $\alpha_s(\mu_H)$, $\alpha_s(\mu_J)$, and
$\alpha_s(\mu_S)$ respectively. $U_H$ and $U_S^\tau$ are evolution kernels which
sum $\alpha_s^i\ln^j\!\ttwoone$ terms. See~\cite{Becher:2008cf} for details.  

The natural scale choices are
\begin{align}
 \mu_H & = m_Z , 
 & \mu_J & = \mu_Q\, \sqrt{\ttwoone} ,
 & \mu_S & = \mu_Q\, \ttwoone .
\end{align}
Here $\mu_Q=\widehat\Tau_1 \sqrt{1+Q^2/(2m_Z^2)}$ is an
average over $\theta$ of $\widehat\Tau_1\beta_\theta$ which appears in the large logarithms.
For $Q=0$ one has $\mu_Q=m_Z$, while for $Q\to\infty$ one has
$\mu_Q=m_Z/(2\sqrt{2})$.  We perform the $s_{1,2}$ and $z_s$ integrations in
\eqref{eq:Ufact} analytically and the $\theta$ integral numerically.  

%
%
\begin{figure}[t!]
\includegraphics[width=0.86\columnwidth]{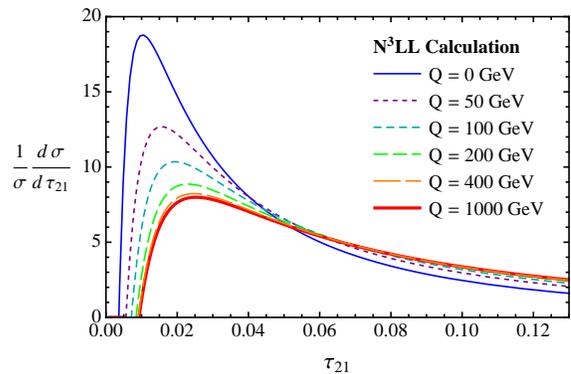}
\vspace{-0.3cm}
\caption{Results of the N$^3$LL analytic calculation for $\ttwoone$ with $\Phi = 0$.  The distribution saturates for $Q \gtrsim 400$ GeV.}
\label{fig:maxAccuracyPlots}
\end{figure}
Results for the $\ttwoone$ distribution for various $Q$ are shown in
\Fig{fig:maxAccuracyPlots}.  As anticipated, the curves rapidly approach a fixed
distribution at large $Q$.
%
%
\begin{figure}[t!]
\includegraphics[width=0.86\columnwidth]{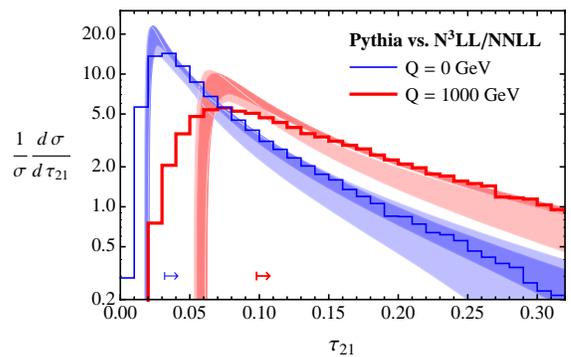}
\vspace{-0.3cm}
\caption{Comparison of theory prediction (bands) for $\ttwoone$ to baseline \pythia (histograms). The heavier (lighter) band is N$^3$LL (NNLL), with widths given by factor of two variations of the hard, jet, and soft scales.  Here, $\Phi = 700$~MeV.  Arrows indicate the approximate range of validity of \eqref{eq:Ufact}.}
\label{fig:vspythia}
\end{figure}

In \Fig{fig:vspythia} we show a comparison to a ``baseline'' \pythia
distribution, where the effects of hadronization are included but the $Z$
width, finite cone size, and ISR/UE contamination have been turned off.  For this comparison we fix $\Phi=700~{\rm MeV}$ to
match the peak of the $Q=0$ \pythia distribution, which allows us to compute the
distribution for all $Q\neq 0$.  In the tail of the distribution, there is
excellent quantitative agreement. The accuracy of \pythia's tail is somewhat artificial since it was tuned to closely related $e^+e^-$ thrust data at $Q=0$.  Predictions in the peak region require additional nonperturbative corrections, which could be included following~\cite{Abbate:2010xh}.

\begin{figure}[t!]
%
%
\subfigure{
\raisebox{4.5cm}{a)\hspace{-0.3cm}} 
\includegraphics[width=0.85\columnwidth]{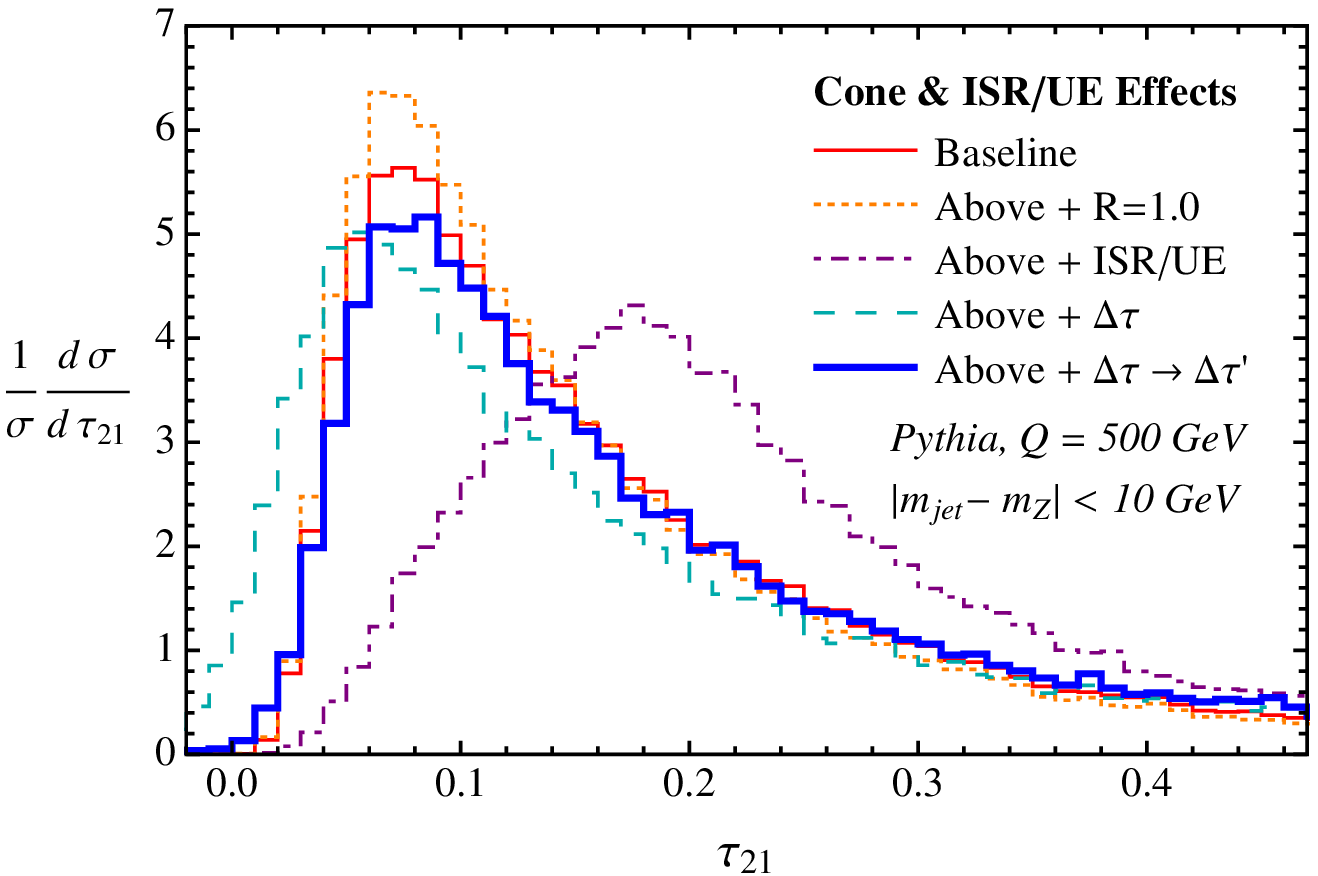}
 \label{eq:Pythia}
}
%
%
\subfigure{
\raisebox{4.5cm}{b)\hspace{-0.3cm}}
\includegraphics[width=0.85\columnwidth]{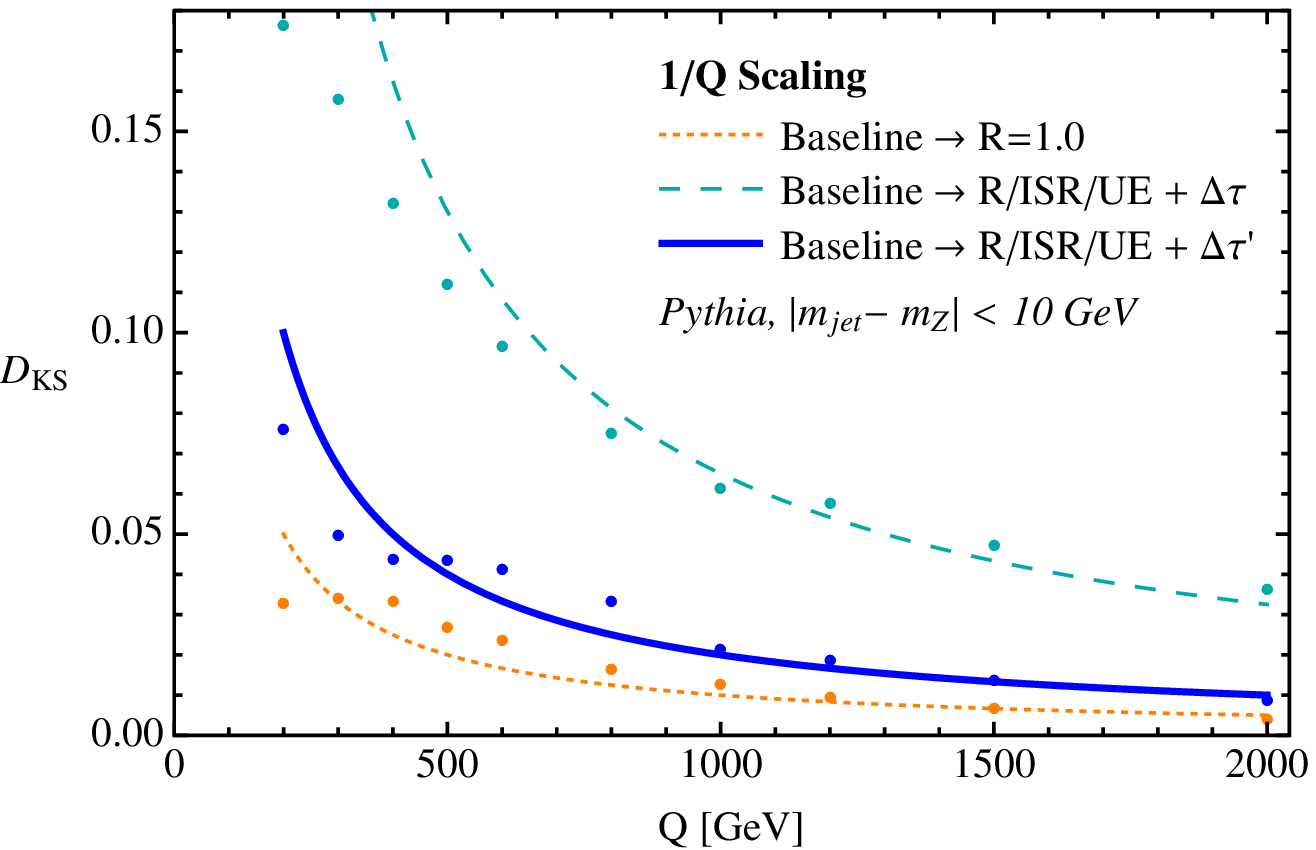}
\label{fig:Qscaling}
}
%
%
%
\subfigure{
\raisebox{4.5cm}{c)\hspace{-0.3cm}}
\includegraphics[width=0.85\columnwidth]{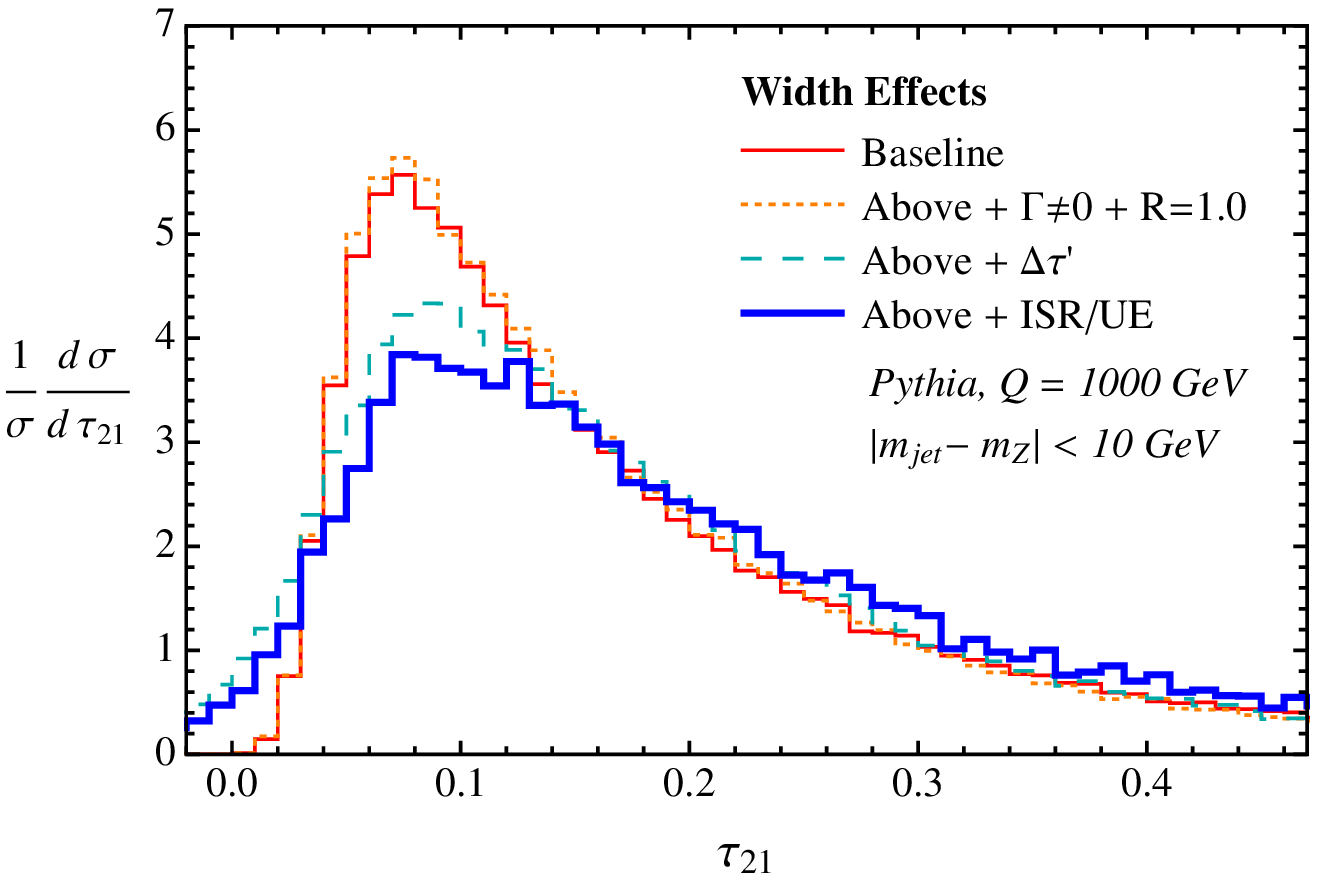}
\label{fig:Finitewidth}
}
%
%
%
\subfigure{
\raisebox{4.5cm}{d)\hspace{-0.3cm}}
\includegraphics[width=0.85\columnwidth]{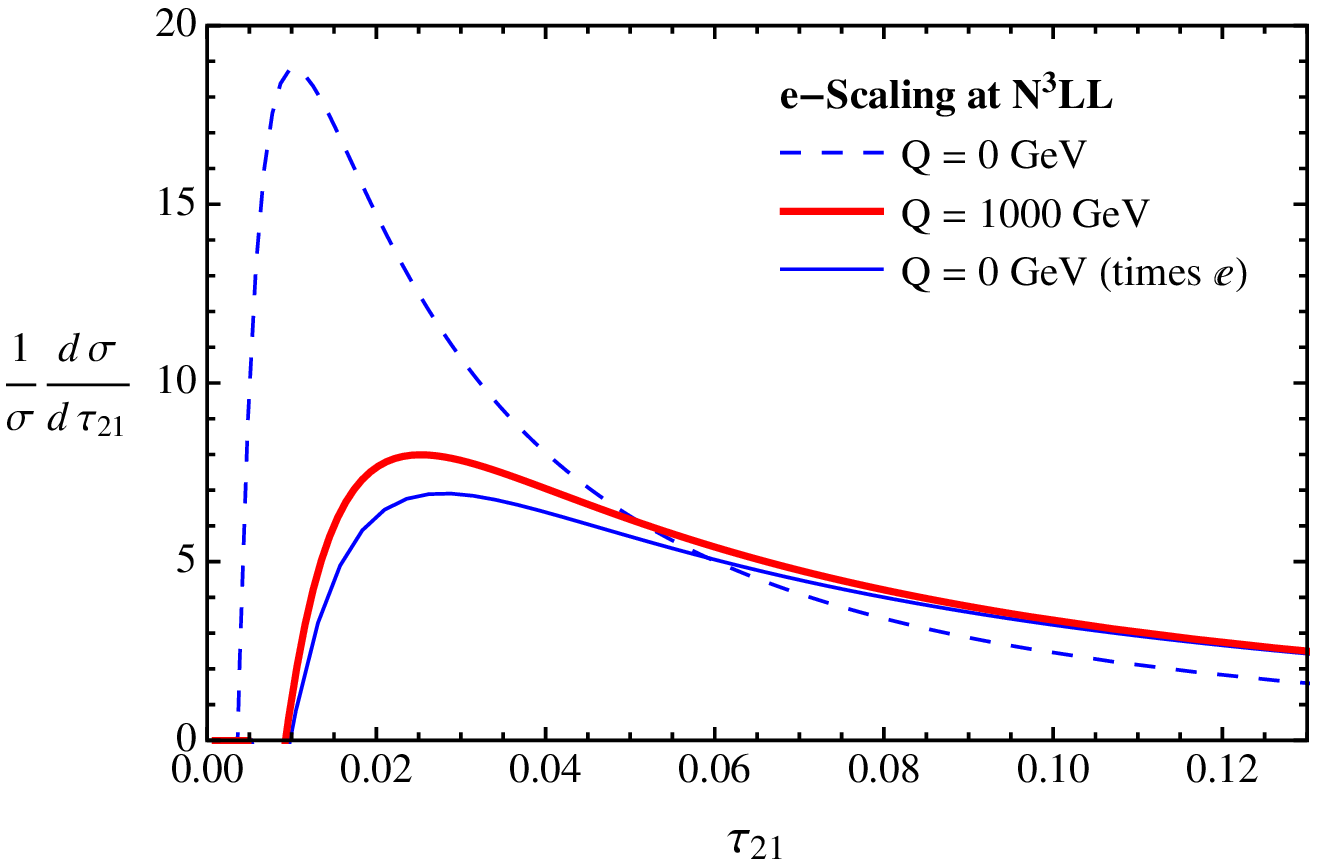}
\label{fig:Escaling}
}
\vspace{-0.4cm}
%
\caption{a) Effect of finite jet cone and ISR/UE in \pythia. The $\Delta \tau'$
  correction mitigates ISR/UE jet contamination. b) Fractional effect of adding
  finite cone and ISR/UE to the \pythia\ baseline distribution.  With the
  $\Delta \tau'$ correction, these effects are smaller than 5\% for $Q \gtrsim
  400$ GeV, and scale as $1/Q$ as expected. c) Effect of finite $Z$ width. d)
  $e$-scaling between $Q = 0$ (thrust) and $Q = \infty$.
}
\end{figure}

In \Fig{eq:Pythia}, we show the effect of a finite $R = 1.0$ cone and jet
contamination in \pythia, restricting our attention to jets whose mass is within
a 10 GeV window of $m_Z$.  At large $Q$, the effect of an $R= 1.0$ cone is quite
mild.  While ISR/UE give a large distortion to $\Tau_2/\Tau_1$, this is
successfully corrected in $\ttwoone$ by the $\Delta \tau$ in \eqref{eq:T12}.
With the $\Delta \tau \rightarrow \Delta \tau'$ replacement we do even better.
Using $\Delta\tau'$ for $Q=1000\,{\rm GeV}$, the \pythia $\ttwoone$ distribution
with $R=1.0$/ISR/UE is indistinguishable at the 2\% level from the baseline distribution shown in
\Fig{fig:vspythia}. Thus our analytic result agrees very well with the full \pythia
distribution.

We use \pythia\ to verify that the effects we have neglected in our calculation
are indeed $1/Q$ suppressed.  In \Fig{fig:Qscaling}, we plot the
Kolmogorov-Smirnov $D$-statistic between the baseline \pythia\ distribution and
\pythia\ as finite cone and ISR/UE effects are reinstated, as a function of $Q$.
The $D$-statistic measures the maximum fractional difference between the
cumulant $\ttwoone$ distributions. Both finite cone and ISR/UE effects fall
off as $1/Q$, and the corrections are $\lesssim 5\%$ for $Q \gtrsim 400$ GeV.

In the above calculation, we neglected the finite width of the $Z$ boson, which
leads to $\mathcal{O}(\Gamma_Z/m_Z)$ corrections that are independent of $Q$.
As shown in \Fig{fig:Finitewidth}, finite width has only a small effect on the
baseline distribution.  Including $\Delta \tau$ yields a larger effect, since
\eqref{eq:deltaTau} assumed that all deviations from the $Z$ pole were due to
jet contamination and not $\Gamma_Z$. Nevertheless, we see in
\Fig{fig:Finitewidth} that $\Delta \tau'$ still mitigates the effect of ISR/UE.
Though beyond the scope of this letter, one can directly
calculate $\ttwoone$ with finite width effects.

It is interesting to explore analytically the $Q$ dependence of
our $d\sigma/d\ttwoone$ (dropping cone and ISR/UE effects and taking $\Phi = 0$) by considering two
extreme cases.  In the $Z$ rest frame $Q =0$, $d\sigma/d\ttwoone$ is equal to
thrust $d\sigma/d\tau$.  In the $Q\to \infty$ limit, $d\sigma/d\ttwoone$ depends
logarithmically on $\ttwoone$ multiplied by various functions of the helicity
angle $\theta$.  Isotropically averaging over $\theta$, these logarithms behave
as
\be
 \int\! \frac{d\cos\theta}{2}  \,  
 \log^n\!\left(\! \tau \sin^2 \frac{\theta}{2} \right) =  \log^n \frac{\tau}{e} + \mathcal{O}\big(\log^{n-2} \tau\big).
\ee
Thus, up to NLL order, the $Q\to \infty$ distribution is related to thrust by scaling by a factor of $e =
2.718...$,
\begin{align}
 \frac{d\sigma}{d\ttwoone}\bigg|_{Q\to\infty} 
  = \frac{1}{e} \frac{d\sigma}{d\tau} (\tau=\ttwoone/e) \,.
\end{align}
This is demonstrated in \Fig{fig:Escaling}.

Our technique of treating the jet boundary and external radiation as $1/Q$
corrections can be readily generalized to color neutral objects with $N$-prong
decays, and the known NNLL ingredients for the $N$-jettiness event
shape~\cite{Jouttenus:2011wh} are a starting point for the calculation of
$N$-subjettiness. It can also be used to compute the distribution of individual
subjet masses $m_i$, which are directly accessible with the $N$-jettiness
factorization theorem. Another straightforward generalization would be to
incorporate massive final state quarks as in $H\to b\bar b$. To treat colored
objects like boosted top quarks (or to calculate the QCD background from light
quark and gluon jets) requires understanding the effect of final-state radiation 
on substructure observables, and we anticipate that expanding about the $Q
\to \infty$ limit will be fruitful in that context as well.

\begin{acknowledgments}
  We thank M.~Baumgart and F.~Tackmann for collaboration at an early stage of
  this work. This work was supported by the U.S. Department of Energy (DOE)
  under the contracts DE-FG02-94ER40818, DE-FG02-05ER-41360, DE-FG02-11ER-41741,
  and DE-SC003916. I.F. is supported by NSERC of Canada.
\end{acknowledgments}

\bibliography{TwoSubjetty}

\end{document}